\documentclass[twocolumn]{aastex63}

\received{April 2020}
\submitjournal{ApJL}

\shorttitle{Runaway gas cooling in a high-redshift cluster}
\shortauthors{Hlavacek-Larrondo et al.}
\begin{document}

\title{Evidence of runaway gas cooling in the absence of supermassive black hole feedback at the epoch of cluster formation}

\correspondingauthor{Julie Hlavacek-Larrondo}
\email{juliehl@astro.umontreal.ca}

\author[0000-0001-7271-7340]{J. Hlavacek-Larrondo}
\affiliation{Département de Physique, Université de Montréal, Succ. Centre-Ville, Montréal, Québec, H3C 3J7, Canada}

\author[0000-0003-2001-1076]{C.L. Rhea}
\affiliation{Département de Physique, Université de Montréal, Succ. Centre-Ville, Montréal, Québec, H3C 3J7, Canada}

\author{T. Webb}
\affiliation{Department of Physics, McGill University, Montréal, Québec, CA}

\author{M. McDonald}
\affiliation{Kavli Institute for Astrophysics and Space Research, MIT, 77
Massachusetts Avenue, Cambridge, MA 02139, USA}

\author{A. Muzzin}
\affiliation{York University, 4700 Keele Street, Toronto, ON, M3J 1P3, Canada}

\author{G. Wilson}
\affiliation{Department of Physics and Astronomy, University of California Riverside, 900 University Avenue,
Riverside, CA 92521, USA}

\author{K. Finner}
\affiliation{Yonsei University, Department of Astronomy, Seoul, Korea}

\author{F. Valin}
\affiliation{Department of Physics, McGill University, Montréal, Québec, CA}

\author{N. Bonaventura}
\affiliation{Niels Bohr Institute, University of Copenhagen, Copenhagen 172100}

\author{M. Cooper}
\affiliation{Center for Cosmology, Department of Physics $\&$ Astronomy, 4129 Reines Hall, University of California, Irvine, CA 92697, USA}

\author{A. C. Fabian}
\affiliation{Institute of Astronomy, Madingley Road, Cambridge CB3 0HA}

\author{M.-L. Gendron-Marsolais}
\affiliation{European Southern Observatory, Alonso de Cordova 3107, Vitacura, Casilla 19001, Santiago, Chile}

\author{M. J. Jee}
\affiliation{Yonsei University, Department of Astronomy, Seoul, Korea}
\affiliation{Department of Physics, University of California, Davis, California, USA}

\author{C. Lidman}
\affiliation{The Research School of Astronomy and Astrophysics, Australian National University, ACT 2601,
Australia}

\author{M. Mezcua}
\affiliation{Institute of Space Sciences (ICE, CSCIC), Campus UAB, Carrer de Can Magrans, 08193, Barcelona,
Spain}
\affiliation{Institut d’Estudis Espacials de Catalunya (IEEC), C/ Gran Capita, 08034 Barcelona, Spain}

\author{A. Noble}
\affiliation{Arizona State University, School of Earth and Space Exploration, Tempe, AZ 871404, USA}

\author{H. R. Russell}
\affiliation{Institute of Astronomy, Madingley Road, Cambridge CB3 0HA}
\affiliation{School of Physics and Astronomy, University of Nottingham, Nottingham NG7 2RD, UK}

\author{J. Surace}
\affiliation{Infrared Processing $\&$ Analysis Center, MS 100-22, California Institute of Technology, Pasadena, CA, 91125, USA}

\author{A. Trudeau}
\affiliation{Department of Physics $\&$ Astronomy, University of Victoria, 3800 Finnerty Road, Victoria, British
Columbia, V8W 2Y2, Canada}

\author{H. K. C. Yee}
\affiliation{Department of Astronomy and Astrophysics, University of Toronto, 50 St George Street, Toronto,
Ontario M5S 3H4, Canada}



\begin{abstract}
Cosmological simulations, as well as mounting evidence from observations, have shown that supermassive black holes play a fundamental role in regulating the formation of stars throughout cosmic time. This has been clearly demonstrated in the case of galaxy clusters in which powerful feedback from the central black hole is preventing the hot intracluster gas from cooling catastrophically, thus reducing the expected star formation rates by orders of magnitude. These conclusions have however been almost entirely based on nearby clusters. Based on new Chandra X-ray observations, we present the first observational evidence for massive, runaway cooling occurring in the absence of supermassive black hole feedback in the high-redshift galaxy cluster SpARCS104922.6+564032.5 ($z=1.709$). The hot intracluster gas appears to be fueling a massive burst of star formation ($\approx900$~M$_\odot$yr$^{-1}$) that is offset by dozens of kpc from the central galaxy. The burst is co-spatial with the coolest intracluster gas but not associated with any galaxy in the cluster. In less than 100 million years, such runaway cooling can form the same amount of stars as in the Milky Way. Intracluster stars are therefore not only produced by tidal stripping and the disruption of cluster galaxies, but can also be produced by runaway cooling of hot intracluster gas at early times. Overall, these observations show the dramatic impact when supermassive black hole feedback fails to operate in clusters. They indicate that in the highest overdensities such as clusters and proto-clusters, runaway cooling may be a new and important mechanism for fueling massive bursts of star formation in the early universe.
\end{abstract}
\section{Introduction}
Galaxy clusters are extremely massive structures that
contain hundreds to thousands
of galaxies, a substantial dark matter component and a large
quantity of hot intracluster gas. At extreme temperatures of tens of
millions of degrees, the central density of the hot gas in many clusters is so high,
that it is expected to cool down to temperatures of $\approx$30~K in less
than a few hundred million years (e.g. \citealt{peterson_x-ray_2006}). Once cooled, this gas should
deposit itself onto the central dominant galaxy, known as the
brightest cluster galaxy (BCG), and extreme star formation rates (SFRs) of
hundreds to thousands of solar masses per year are expected (e.g. \citealt{fabian_cooling_1994}). However, observations have shown that the observed SFRs are orders of magnitude lower. We now understand that it is the supermassive black hole (SMBH) in the BCG
that is preventing the hot intracluster gas from cooling by driving supersonic jets that carve out gigantic X-ray cavities (see a review by \citealt{mcnamara_mechanical_2012}), a process known as mechanical active galactic nucleus (AGN) feedback.

However, most of our understanding of AGN feedback in clusters has been based on nearby
objects and it has remained observationnally challenging
to determine if such feedback is also occurring in distant
clusters (e.g. \citealt{hlavacek-larrondo_extreme_2012}; \citealt{hlavacek-larrondo_x-ray_2015}; \citealt{birzan_study_2017}). This is due to the fact that nearby clusters are more easily
studied given their proximity, but also because of the lack of
well-understood samples of high-redshift clusters. 

The situation has now dramatically
changed with the advent of new cluster surveys. Combined with extensive follow-up Chandra observations, the $2500$~deg$^2$ SPT cluster
survey (\citealt{vanderlinde_galaxy_2010}, \citealt{reichardt_galaxy_2013}) has proven to be
a key player for our understanding of cluster evolution at $z>1$ (e.g. \citealt{rossetti_cool_2017}; \citealt{mcdonald_remarkable_2017}; \citealt{mcdonald_evolution_2016}) and showed that powerful mechanical AGN feedback has been operating in at least some clusters since $z\approx1$; \citealt{hlavacek-larrondo_x-ray_2015}). 

The SpARCS cluster
and  Stellar Bump Sequence surveys have also discovered
over 500 $z>0.6$ clusters in the Spitzer SWIRE fields (e.g. \citealt{wilson_spectroscopic_2009}; \citealt{muzzin_spectroscopic_2009}; \citealt{muzzin_evolution_2013}). \cite{webb_star_2015} showed that beyond z$\approx$1, significant in situ star formation seems to be
occurring at the cores of clusters. This is in direct contrast to what
is seen in the local universe, in which the central AGN is preventing star
formation from occurring. Using the SPT sample, \cite{mcdonald_star-forming_2016} found a similar result. Both studies suggest that beyond
z$\approx$1, the star formation in BCGs may be driven by gas-rich major
mergers instead of residual cooling flows. These conclusions were motivated by the change in slope of the specific SFR (sSFR) with redshift and one case study of an apparent gas-rich BCG merger in the cluster SpARCS104922.6+564032.5 (hereafter SpARCS1049; \citealt{webb_extreme_2015}).

\subsection{SpARCS104922.6+564032.5}
SpARCS1049 was first identified in 2015 as an optically rich system located at $z=1.709$ with 27 spectroscopically confirmed members (\citealt{webb_extreme_2015}). It has a
richness-estimated mass within 500 kpc of $3.8\pm1.2\times10^{14}$M$_\odot$, placing it at an extremely important epoch in which the most
massive structures in the universe were forming. 

A recent weak-lensing
analysis of the cluster based on infrared Hubble Space Telescope (HST)
observations confirms its high-mass of $3.5\pm1.2\times10^{14}$M$_\odot$
and suggests that the cluster has no significant substructure (\citealt{finner_notitle_2020}). The HST observations also revealed an unusual
long ($\approx$60 kpc) tidal-like feature in the core of the cluster that was thought to originate
from a gas-rich major merger given its morphology and that it was found to coincide with an extreme
infrared source ($L_{\rm{IR}}=6.2\pm0.9\times10^{12}L_\odot$; \citealt{webb_extreme_2015}). Spitzer infrared spectrograph observations found that the infrared source was
also coincident with polycyclic aromatic hydrocarbons features
at the redshift of the cluster (\citealt{farrah_high_2007}), indicating that the
emission was dominated by star formation and not from an accreting SMBH. Overall, these observations showed that the cluster core appears to host an extreme starburst with a (AGN-corrected) SFR of $860\pm130~$M$_\odot$yr$^{-1}$ (\citealt{webb_extreme_2015}; \citealt{webb_star_2015}). 

The only other cluster known to host such an extreme starburst at its core is SPT-CL~J2344-4243, i.e. the Phoenix cluster located at $z=0.597$ with a SFR of $500-800~$M$_\odot$yr$^{-1}$ (\citealt{mcdonald_massive_2012} and references therein). In this case, extreme AGN feedback is occurring (as seen from X-ray cavities,
radio jets and a central quasar), but it appears to be insufficient
to suppress cooling of the hot intracluster gas (\citealt{mcdonald_anatomy_2019}).

However, in the case of SpARCS1049, the 24 micron Spitzer MIPS
emission was unusual and appeared to be offset by $\approx$25~kpc from the central galaxy and
not associated with any other cluster member (in direct contrast to the star formation occurring in the Phoenix cluster). Such features may have suggested that the intense star bursting occurring in SpARCS1049 is being driven by a merger-like event, but an extremely large molecular gas reservoir of $1.1\pm0.1\times10^{11}~$M$_\odot$ was also detected in the core (\citealt{webb_detection_2017}) and showed no
signs of multiple velocity peaks as would be expected in a
major-merger event (\citealt{greve_interferometric_2005}; \citealt{gao_molecular_2001}; \citealt{schulz_interstellar_2007}). Recently, these features were also
interpreted as evidence of ram pressure stripping occurring in the
cluster core (\citealt{castignani_environmental_2020}). 

Here, we present the first X-ray observations of SpARCS1049 (PI Hlavacek-Larrondo). We show that X-rays provide a key missing piece of the puzzle: they reveal that the starburst is directly linked to the intracluster gas and consistent with being fueled by massive runaway cooling of a cool core. This is in direct contrast to what is seen in nearby clusters and indicates that runaway cooling may be a new and important mechanism for fueling massive bursts of star formation in the early universe for the highest overdensities. In Section 2, we present the observations. In Section 3, we discuss the results and in Section 4 their implications. Throughout this paper, we assume $H_0=70$~km s$^{-1}$ Mpc$^{-1}$, $\Omega_m=0.3$ and $\Omega_\Lambda=0.7$. All errors are 1$\sigma$ and all energy bands are in the observer's frame unless otherwise specified.

\section{Observations and data reduction}

\subsection{Chandra X-ray Observations}

\begin{figure}[ht]
    \centering
    \includegraphics[width=0.45\textwidth,clip]{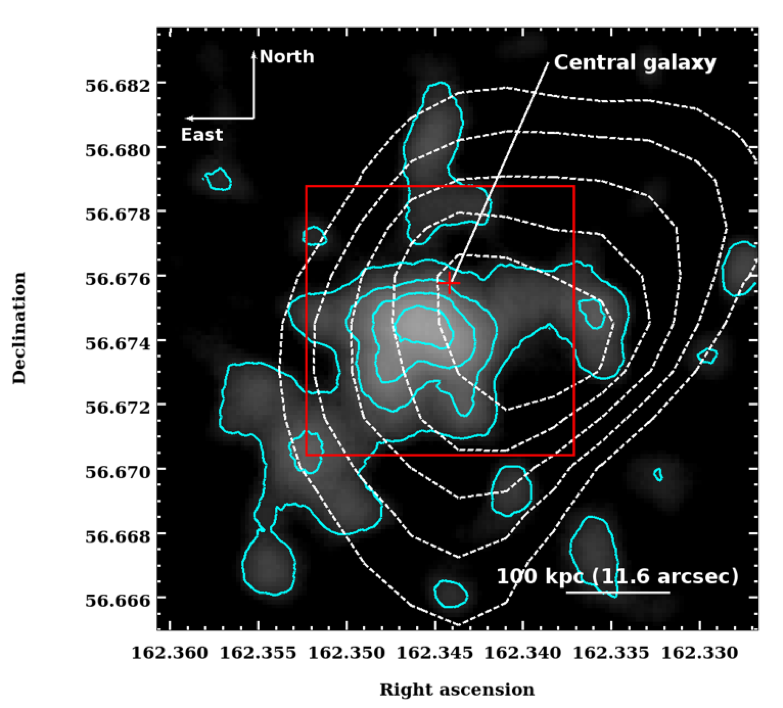}
    \caption{Merged exposure-corrected, background-subtracted $0.5-7.0$ keV Chandra X-ray image of SpARCS1049. The image has not been binned, but smoothed with a Gaussian function of $\sigma=5$ pixels. In cyan, there are 4 contours starting at 4$\sigma_{\rm{rms}}$ where $\sigma_{\rm{rms}}$ is the standard deviation in the background count per pixel located on the same ACIS-I3 chip, but several Mpc outside the cluster. The BCG is also shown with the red cross symbol (see \citealt{webb_extreme_2015} for method of identification). The X-ray emission is highly peaked and consistent with a compact cool core that is offset from the central galaxy. The white dashed contours show the weak lensing mass reconstruction (\citealt{finner_notitle_2020}). The red box is the zoomed-in region of Figure \ref{fig:MultiWavelengthzoom}.}
    \label{fig:MultiWavelength}
\end{figure}

The first X-ray observations of SpARCS1049
were obtained with ACIS onboard the Chandra X-ray Observatory (PI Hlavacek-Larrondo). The object was observed in 2018 for 170 ks (ObsIDs 20528, 20941, 20940 and 21129). All observations were centered on ACIS-I3. The data were reduced using CIAO v4.11. Due to the low counts and extended nature of the object, we did not follow the standard reduction pipeline. Instead, we constructed a level 2 event file while mimicking the steps of several other authors with the goal of maximizing the number of counts of our source (\citealt{broos_innovations_2010}; \citealt{weismann_studying_2013}). After correcting for the initial astrometric alignment, we used the task lc\_sigma\_clip to investigate the presence of major flares, but no event above 3$\sigma$ was detected. We then used destreak to clear the event file of residual streaks. In creating the bad-pixel file, we used a custom bitflag which allowed us to retain more counts in the diffuse regions. In the final step, we executed the acis\_process\_events with check\_vf\_pha set to yes, a process that improves the signal-to-noise ratio for diffuse sources such as SpARCS1049. An exposure map assuming a monoenergetic photon distribution at 1.53~keV, corresponding to the peak expected for a massive cluster at $z\approx1.7$, was used to create the merged, background-subtracted and exposure-corrected image shown in Figure \ref{fig:MultiWavelength}. We note that we also ran the standard reduction pipeline developed by the Chandra X-ray Center. The cluster is detected in both cases and the results of this paper remain unchanged, but our tailored pipeline allows us to maximize the cluster counts.

\subsubsection{Astrometric Corrections}

The HST frames were initially aligned for co-addition using the Drizzle package. Source Extractor was then used to extract sources in the HST images (7 stars were found) and a script was built to match these sources to those in a reference catalog. We use Gaia as the reference catalog and found a systematic offset of the matched stars of RA$=0.5\pm0.1$'' and Dec$=-0.2\pm0.1$''. The offset was corrected by adjusting the WCS of the HST images. The accuracy of $\simeq0.1$'' is determined by the reference frame. We then examined the Chandra X-ray images and found that half a dozen galaxies detected in the HST images had bright X-ray point sources associated with them (presumably from the central AGN). They were all systematically offset by $0.25''$ to the south east. We re-aligned the X-ray images and use these throughout this work.

\subsubsection{Photometric and spectroscopic analysis}\label{meth:SF}

\begin{figure}[ht]
    \centering
    \includegraphics[width=0.45\textwidth]{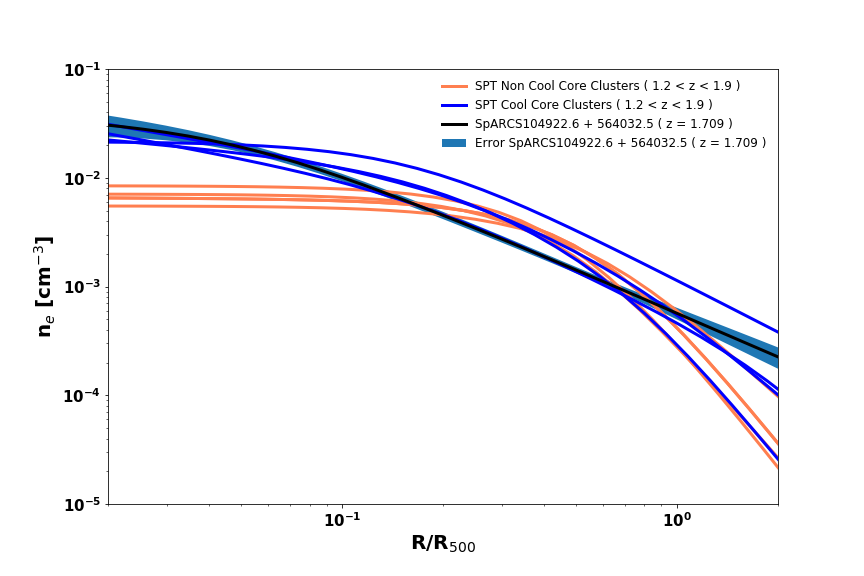}
    \caption{Deprojected electron density profile of SpARCS1049 assuming spherical geometry and scaled for R$_{500}$ (black curve). Profiles of the eight $1.2<z<1.9$ SPT clusters of galaxies that have Chandra X-ray observations are also shown (see \citealt{mcdonald_remarkable_2017} for details). The top 4 are cool core clusters as defined by they central electron density value. This figure shows that SpARCS1049 has an over dense core (i.e. a cool core).}
    \label{fig:Profile}
\end{figure}

The data were spectroscopically fit using Xspec v12.10.1, Sherpa v1, and python v3.5. Given the low count rate, we could not constrain the redshift of the X-ray source and assumed $z=1.709$ (\citealt{webb_extreme_2015}, \citealt{webb_detection_2017}; see also Appendix A for evidence that the X-ray source is indeed associated with the cluster). We also could not constrain the metallicity and assumed a ratio of 0.3 Z$_\odot$ (e.g. \citealt{anders_abundances_1989}; \citealt{arnaud_iron_1992}; \citealt{bulbul_high-resolution_2012}; \citealt{molendi_critical_2016}; \citealt{mcdonald_evolution_2016}). Note that we re-derived all quantities using a ratio of 0.2 Z$_\odot$ and found consistent results. To account for Galactic absorption, we used a fixed value of $5.99\times10^{19}\text{cm}^{-2}$ for the hydrogen column density (\citealt{kalberla_leidenargentinebonn_2005}). The background region was chosen to be on the ACIS-I0 chip at several Mpc from the cluster. We also considered a blank sky background and found consistent results. We fit the source and background regions of each observation simultaneously. We modeled the background emission following the methods of \cite{sun_chandra_2009} and \cite{mcdonald_remarkable_2017}. Both methods replicate the soft and hard excesses observed in the cosmic X-ray background. We found consistent results and opted to use the McDonald model so that we can directly compare our results with theirs. This model includes a soft X-ray Galactic component ({\sc{apec}}, $kT=0.18$keV, Z$=0$) and a hard cosmic X-ray component ({\sc{bremss}}, $kT=40$keV). To account for the cluster emission, we considered the {\sc{apec}} and {\sc{mekal}} models and found consistent results. All values quoted hereafter have been derived with {\sc{apec}}. Since we are in a low-count regime, we also use c-statistic and conduct all fits using the single energy range $0.6-5.0$~keV. We find that the target has an integrated rest-frame $2-10$ keV X-ray luminosity of $4.29\pm0.19\times10^{44}$ erg/s (equivalent to a flux of $2.18\pm0.10\times10^{-14}$ erg/s/cm$
^{-2}$) and a temperature of $5.71\pm1.57$ keV within 200 kpc of the peak X-ray emission.
The overall morphology is compact and reminiscent of a relaxed galaxy cluster with a mild elongation in the north-west to south-east direction. The X-ray surface brightness concentration (C$_{\rm{SB}}$ = $0.19_{-0.05}^{+0.07}$), defined as the ratio between the energy flux within 40 kpc and within 400 kpc in the 0.5 to 2.0 keV band, indicates that SpARCS1049 has an overdense core (i.e. a cool core).  It is one of the few known clusters with an overdense core at $z>1.5$ (e.g. \citealt{mcdonald_remarkable_2017}). In Figure \ref{fig:Profile}, we show the deprojected density profile as a function of radius normalized by R$_{500}$, compared to the high-redshift SPT clusters. Profiles were determined following the methods of \cite{vikhlinin_chandra_2006}, \cite{andersson_x-ray_2011}, and \cite{mcdonald_growth_2013}. We refer the reader to these papers for the details. We also show the 1$\sigma$ uncertainties on the profile, determined by running the fits 100 times while bootstrapping the uncertainties. Note that the profile does not include any uncertainty in the temperature since the method assumes a temperature profile when converting from emission measure to density. Following this method, we find R$_{500}$ to be $\approx$450~kpc for SpARCS1049. This value is consistent with the expected R$_{500}$ value from the weak-lensing mass estimate ($\approx$600~kpc). Figure \ref{fig:Profile} shows that the deprojected density profile is highly peaked with a central density ($n_{e,0}=0.07\text{cm}^{-3}$) that is again indicative of a cool core. Here, central electron density divides cool cores and non cool cores at n$_{e,0}=0.015$ (e.g. \citealt{hudson_what_2010}; \citealt{mcdonald_growth_2013}). We note that the profile beyond $\approx50$kpc follows a different slope compared to other clusters; implying that the outer parts of the cluster, usually driven by self-similar processes, may not yet be well established in this cluster. Cool cores may therefore form before self-similar processes are established in the outer regions of clusters (e.g. \citealt{vikhlinin_chandra_2006}; \citealt{croston_galaxy-cluster_2008}; \citealt{mantz_cosmology_2015}). Another possibility, as we will explore in more detail in Section \ref{Sec3.2}, is that the cluster may have recently undergone a merger that has displaced the cool core from the center of the potential. The outer parts of the cluster may therefore still reflect this merger.

\begin{figure}[ht]
    \centering
    \includegraphics[width=0.45\textwidth,clip]{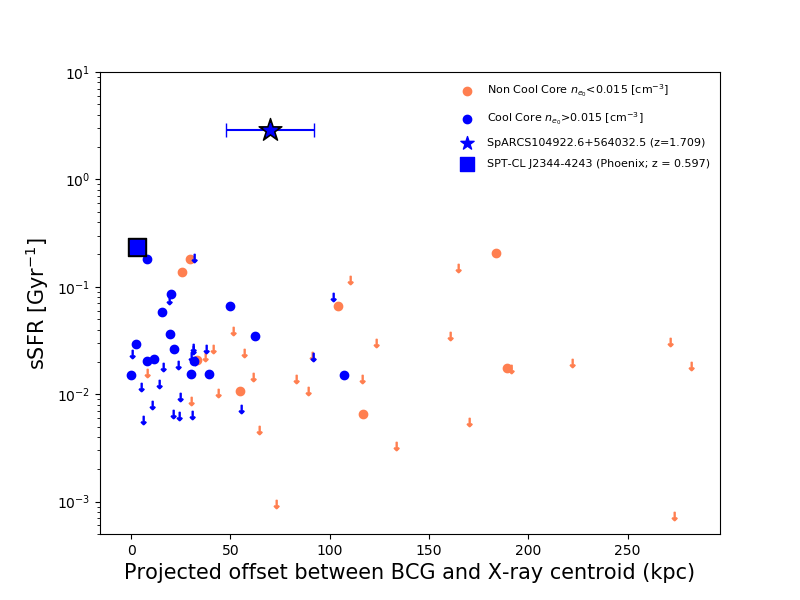}
    \caption{Comparison of the star formation processes occurring in SpARCS1049 to those occurring in the SPT galaxy clusters. The plot shows the sSFR as a function of the projected distance between the BCG and the centroid of the cluster X-ray emission. The clusters are color-coded depending on the central value of their deprojected electron density profile.}
    \label{fig:SFR}
\end{figure}

Figure \ref{fig:SFR} compares the sSFR in SpARCS1049 to the SPT clusters. Here, we extracted the redshifts and positions of the BCGs from \cite{mcdonald_evolution_2016}. X-ray centroids were taken from \cite{mcdonald_growth_2013}. With these positions, we used astropy's separation function to calculate the projected offset between these two quantities. We then extracted all the available values for the SFR of the BCGs (UV, O[II], 24 microns). For systems detected in one or more bands, the average of the detected SFRs was used to represent the SFR, disregarding any upper limits. For sources with only upper limits, we calculated the average and treated this as an upper limit. To calculate the sSFR, we divided the SFR by the BCG stellar mass. We further subdivided the clusters according to the value of their central deprojected electron density (n$_{e,0}$). For SpARCS1049, the SFR and BCG stellar mass were taken from \cite{webb_extreme_2015}. The X-ray centroid was determined using the iterative procedure of \cite{cavagnolo_entropy_2009}, and includes a statistical error based on the ciao tools. 
\subsection{VLA Observations}
\begin{figure}[ht]
    \centering
    \includegraphics[width=0.45\textwidth,clip]{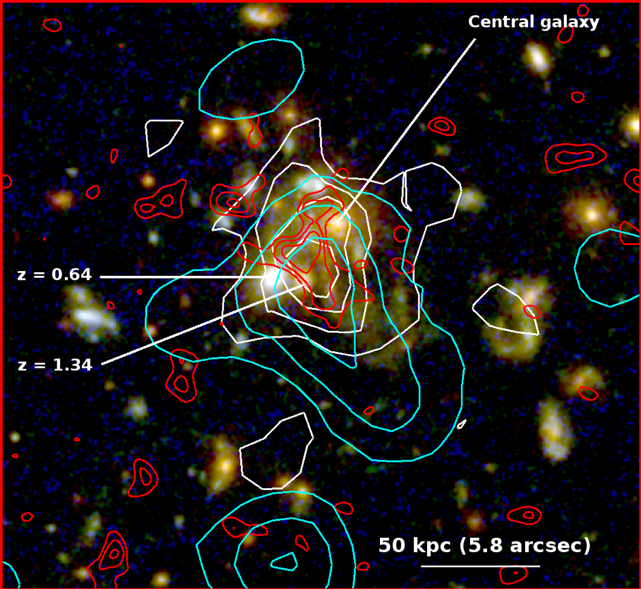}
    \caption{HST color-composite image of the cluster core using the F160W,
        F150W and F814W filters (same region as the red square in Figure \ref{fig:MultiWavelength}). The cyan contours show the $0.7-1.0$ keV X-ray
        emission of the cluster used to highlight the coolest X-ray
        gas that Chandra can detect, starting at
        4$\sigma_{\rm{rms}}$}. We show the new VLA CO (1-0) emission with the red
        contours and Spitzer MIPS 24 micron emission with white
        contours (Valin et al. in prep and \citealt{webb_extreme_2015}). The MIPS emission
        peaks on the tidal-like feature seen in the HST images and
        is slightly extended along the direction of this
        tail. The MIPS emission is also consistent with the location
        of the coolest X-ray gas. This image shows that the location
        of the coolest intracluster gas is co-spatial with the
        location of the star formation in SpARCS1049.
    \label{fig:MultiWavelengthzoom}
\end{figure}

New Q-band observations with the the Very Large Array (VLA) were obtained in 2019 for SpARCS1049 (18B-177; PI Webb). These observations probe the redshifted CO(1-0) line. We briefly summarize the data reduction procedure (the details will be presented in Valin et al. in prep). The C-configuration was chosen to maximize the detection, while allowing for high enough spatial resolution to resolve the molecular gas (beam of $\approx0.47''\approx4~$kpc). The observations were completed in optimal conditions and the data were reduced following the standard CASA procedure (v5.4.2-5). Figure \ref{fig:MultiWavelengthzoom} presents the resulting continuum image ranging from 42.456 GHz to 42.616 GHz obtained with tclean. Contours start at $2\sigma_{\rm{rms}}$, where $\sigma_{\rm{rms}}=45.7$ $\mu$Jy/beam.
\section{Discussion}

\subsection{Runaway gas cooling as the source of the starburst}

In Figure \ref{fig:MultiWavelengthzoom}, we show contours highlighting the coolest intracluster gas detectable with Chandra ($0.7-1.0$ keV) starting at $4\sigma_{\rm{rms}}$, where $\sigma_{\rm{rms}}$ is the standard deviation in the background count per pixel, located on the same ACIS-I3 chip several Mpc outside the cluster. This gas is located near the peak of the X-ray emission and is remarkably co-spatial with the large, $\approx60$ kpc tidal-like feature seen in the HST images (\citealt{webb_extreme_2015}). 

The tidal-like feature was initially thought to originate from a gas-rich major merger occurring in the cluster core, given its morphology and that it was found to coincide with the starburst ($860\pm130~M_\odot$~yr$^{-1}$; \citealt{webb_extreme_2015}; \citealt{webb_star_2015}). It was unusual given that the 24 micron MIPS centroid was significantly offset ($\approx25$ kpc) from the BCG or any other cluster member (see Figure \ref{fig:MultiWavelengthzoom}), indicating that the intense star formation was not associated with any galaxy. The narrow spectral signature of the large molecular gas reservoir in the cluster core ($1.1\pm0.1\times10^{11}$~M$_\odot$; \citealt{webb_detection_2017}) was also in direct contrast to what is expected from a major merger (e.g. \citealt{greve_interferometric_2005}; \citealt{gao_molecular_2001}; \citealt{schulz_interstellar_2007}). Instead, the velocity dispersion of the molecular gas matches the dispersion seen in nearby clusters of galaxies in which a small fraction of the intracluster gas is cooling (\citealt{mcnamara_10_2014}; \citealt{russell_massive_2014};  \citealt{gonzalez_intracluster_2005}).

Recently, \citealt{castignani_environmental_2020} obtained NOEMA CO($4\rightarrow3$) and continuum map observations of SpARCS1049. They detected two sources within 20kpc of the BCG: the first appeared to be associated with a pair of merging cluster galaxies, while the second showed evidence of a CO($4\rightarrow3$) tail and was interpreted as evidence for ram pressure stripping. The authors argued that such mergers in the core could be the source of the starburst.

Here, by imaging the cluster for the first time at X-ray wavelengths, we find that the cluster harbours a cool core and that the coolest intracluster gas is directly co-spatial with the HST tidal-like feature. Our VLA observations also show that the HST tidal-like feature and cool X-ray gas are co-spatial with the CO (1-0) gas. This co-spatiality indicates that the intense infrared source, HST tidal-like feature and molecular gas reservoir must be intimately linked to the hot intracluster gas. 

We do not expect any merger event or ram pressure stripping event to also contain cool X-ray gas associated with the star formation. Indeed, nearby clusters have shown that ram pressure stripping could lead to tails of cool ($\approx0.5-1.0$ keV) X-ray gas trailing behind galaxies, but such X-ray tails have typical $0.5-2.0$ keV X-ray luminosities of $\approx10^{40}$ erg/s. This is 4 orders of magnitude smaller than what is expected for a cool core in a massive cluster - and would be undetectable with 170 ks of Chandra observations at the redshift of SpARCS1049 (e.g. \citealt{zhang_narrow_2013}; \citealt{sun_spectacular_2010}; \citealt{kraft_stripped_2017}). The brightest ram pressure stripped X-ray tail discovered to date has a $0.5-2.0$ keV X-ray luminosity of $10^{42}$ erg/s (e.g. \citealt{schellenberger_long_2015}), which remains too faint to be significantly detected with $Chandra$ in 170 ks at the redshift of SpARCS1049. We also note that the X-ray temperatures of ram pressure stripped X-ray tails (typically $\approx0.5-1.0$ keV) would fall out of the energy range detectable with $Chandra$ once redshifted to $z=1.709$. Overall, this implies that the cool X-ray gas seen in Figure \ref{fig:MultiWavelengthzoom} can not be due to ram pressure stripping.

Instead, our results provide evidence that the intense starburst occurring in the cluster core is likely being fueled by massive, runaway cooling of the intracluster gas. Only a moderate cool core would be required to fuel a starburst of $\approx900M_\odot$~yr$^{-1}$ if allowed to cool completely (e.g. \citealt{fabian_cooling_1994}). 

At low-redshift, the SFR seen in the cores of cool core clusters are typically 1$\%$ of the expected rates (\citealt{peterson_x-ray_2006}; \citealt{odea_infrared_2008}; \citealt{mcdonald_revisiting_2018}). This is because the SMBH in the BCG is preventing these hot halos from cooling via powerful mechanical jetted outflows that inflate large X-ray cavities (\citealt{mcnamara_mechanical_2012}; \citealt{fabian_observational_2012}) and allow only a small fraction of residual cooling to occur. In the case of SpARCS1049, there is no evidence that the SMBH in the BCG is actively accreting : the central galaxy is barely radio-detected and shows no evidence of jetted outflows (\citealt{trudeau_multiwavelength_2019});  we find no evidence of an X-ray point source (indicating the presence of an accreting SMBH) coincident with the central galaxy; and the optical and infrared photometry of the central galaxy shows that it is quiescent. These observations are therefore consistent with runaway cooling of the hot halo occurring because of the absence of feedback from the central galaxy. 

\subsection{A cool core offset from its central galaxy}\label{Sec3.2}

In Section \ref{meth:SF}, we showed that both the X-ray surface brightness coefficient and deprojected central electron density place the cluster in the cool core category. 

The X-ray emission is however slightly elongated in the south-east to north-west direction (see Figure \ref{fig:MultiWavelength}), indicating that the cluster may be undergoing a merger that has not destroyed the cool core (see e.g. A2146 for an example of a cool core that survived a bullet-like merger; \citealt{russell_shock_2012}). A merger scenario could also explain the large $\approx25$ kpc offset between the coolest intracluster gas and BCG, as well as the large $\approx50$ kpc offset between the X-ray peak and BCG (Figures \ref{fig:MultiWavelength} and \ref{fig:MultiWavelengthzoom}). In nearby clusters of galaxies, such large offsets are usually associated with major mergers (\citealt{hudson_what_2010}; \citealt{rossetti_cool_2017}). 

\cite{hamer_relation_2012} identified 3 systems (out of 77 nearby line-emitting BCGs) in which the soft X-ray peak was displaced from the BCG. The soft X-ray peak was also coincident with optical line emission, similar to what is occurring in SpARCS1049 although of very different scales. Indeed, the observed offset between the BCG and the peak H$_\alpha$ emission was at most $\approx$10~kpc versus $\approx25$~kpc in SpARCS1049. The authors argued that such displacements may be caused by a large event such as a major merger (see also \citealt{pasini_bcg_2019} and \citealt{vantyghem_enormous_2019}). Nonetheless, it is important to remember that SpARCS1049 is located at the epoch in which the most massive structures (such as SpARCS1049) are still forming and have not yet settled into equilibrium. It is therefore unclear if such direct comparisons to nearby clusters can be applied to clusters located at $z\approx1.7$, especially given the low number of counts detected in the case of SpARCS1049. Detailed simulations are required to determine if such displacements (and cool core survival) are possible in cluster mergers at the epoch of cluster formation.


\section{Implications}

\subsection{The failure of AGN feedback}

In nearby clusters of galaxies, it has been argued that AGN feedback might form a self-regulated loop in which the jetted outflows trigger instabilities which allow a fraction of the hot gas to cool and rain down onto the central SMBH, re-starting the feedback loop (e.g. \citealt{gaspari_chaotic_2013}; \citealt{voit_regulation_2015}; \citealt{tremblay_cold_2016}). This is consistent with the fact that at low-redshifts, the cool cores are usually centered on the BCGs and that these can directly fuel the central SMBH.

In SpARCS1049, the coolest X-ray gas (and starburst) is offset by $\approx25$~kpc from the BCG and the X-ray peak is offset by almost 50~kpc from the BCG. Given this displacement, the absence of feedback in SpARCS1049 might therefore be caused by a lack of gas supply onto the central SMBH. If gas can not be funnelled down to the central SMBH, then it implies that the central SMBH may not be accreting enough material to power a jet, let alone a jet powerful enough to offset cooling of a cool core. This is in agreement with the recent study of \cite{trudeau_multiwavelength_2019} that found no evidence of radio jets associated with the BCG in SpARCS1049. If this is the reason why runaway gas cooling is occurring in SpARCS1049, our results imply that the self-regulated feedback loop $requires$ cool cores to be spatially aligned with the BCG. It also implies that the central SMBH must be directly fueled by the hot halos.

\subsection{Star formation in high-redshift clusters and proto-clusters}

Figure $\ref{fig:SFR}$ shows that the star formation occurring in the core of SpARCS1049 is orders of magnitude higher than what is seen at low-redshifts. Given that SpARCS1049 is located at the epoch of cluster formation, our results indicate that runaway cooling of intracluster gas can be an important process of star formation in the highest overdensities (i.e. clusters and proto-clusters) at high-redshift. At the very least, our results imply that some of the intense star formation occurring in newly identified clusters and proto-clusters at high-redshift (e.g. \citealt{capak_massive_2011}; \citealt{chiang_galaxy_2017}) may be driven by runaway gas cooling as opposed to galaxy merger processes.

\subsection{A new mechanism for building intracluster stars}

Our results show that runaway cooling can deposit a tremendous amount of newly formed stars in the cores of clusters. In fact, in less than 100 million years, this cooling can form the same amount of stars as in the Milky Way. Consequently, our results directly imply that intracluster stars are not only produced by tidal stripping and the disruption of cluster galaxies (\citealt{gregg_galaxy_1998}; \citealt{conroy_hierarchical_2007}), but can also be produced early on in the cluster life through massive cooling of the intracluster gas. This is consistent with recent studies suggesting the these stars are already in place at $z>1$ (e.g. \citealt{ko_evidence_2018}), implying that runaway cooling of the hot halos can account for part of the intracluster light in clusters (\citealt{lin_k-band_2004}; \citealt{conroy_hierarchical_2007}). Another consequence is that this process appears to be capable of depositing the newly formed stars over dozens of kpc, i.e. the entire cluster core. Runaway gas cooling can therefore easily distribute intracluster stars over large distances. 

\section{Concluding remarks}
Overall, our results directly illustrate the fate of hot X-ray halos when SMBH feedback fails to operate, a process thought to be commonly occurring at cosmic dawn when galaxies were first forming (e.g. \citealt{cattaneo_role_2009}). They directly imply that star formation processes in the early Universe may not only be driven by the classical merger and disc scenarios, but may also be driven by runaway gas cooling in the highest overdensities. 

\acknowledgments
J. H.-L. acknowledges support from NSERC via the Discovery grant program, as well as the Canada Research Chair program. C. R. acknowledges financial support from the physics department of the Universit\'e de Montr\'eal. GW acknowledges support from the National Science Foundation through grant AST-1517863, by HST program number GO-15294, and by grant number 80NSSC17K0019 issued through the NASA Astrophysics Data Analysis Program (ADAP). Support for program number GO-15294 was provided by NASA through a grant from the Space Telescope Science Institute, which is operated by the Association of Universities for Research in Astronomy, Incorporated, under NASA contract NAS5-26555. MJJ acknowledges support for the current research from the National Research Foundation of Korea under the programs 2017R1A2B2004644 and 2020R1A4A2002885. HRR acknowledges support from an STFC Ernest Rutherford Fellowship and an Anne McLaren Fellowship. We also greatly thank the anonymous referees that provided the first reports when submitted to the initial journal, as well as the anonymous referee in the final journal. AT is supported by the NSERC
Postgraduate Scholarship-Doctoral Program.

\newpage
\appendix

\section{Origin of the X-ray emission}

The X-ray source detected at the location of SpARCS1049 has a diameter of $\approx$50'' ($\approx400$ kpc at $z=1.709$) as traced by the 4$\sigma_{\rm{rms}}$ contours (see Fig. \ref{fig:MultiWavelength}), entirely consistent with the X-ray emission originating from a $\approx10^{14}M_\odot$ cluster located at $z\approx1.7$. We detect over 140 X-ray counts (above the background) associated with the object in the $0.5-7.0$ keV energy range. The X-ray luminosity and temperature of the source also fall right along the scaling relations expected for galaxy clusters (e.g. \citealt{anderson_unifying_2015}). The X-ray source cannot be X-ray emission originating from a population of X-ray binaries in the starbursting core as this emission would be two orders of magnitude lower for typical X-ray luminosity to star formation ratios, even in low-metallicity environments. If the X-ray emission originated from a background source, then the only structure that could explain the large X-ray luminosity of $\geq10^{44}$ erg s$^{-1}$, extended morphology and high temperature would be another massive cluster located at  $z>>1.7$. It is statistically unlikely to have two large over densities overlap each other within such a small region. In addition, Webb et al. (2015) carried out a campaign of near-infrared (NIR) spectroscopy with MOSFIRE on Keck on the field in which SpARCS1049 is located. This was combined with a literature search for redshifts from other instruments. It was found that the most massive structure at $z\approx1.3-2.0$ is the $z=1.709$ cluster. Since then, new GMOS Gemini observations were obtained (PI Webb). These new grating observations trace the full redshift range in a single mask at $0.3<z<1.7$ within $2.5'$ of the X-ray detection. We probed the [OII] emission over the redshift range $0.3<z<1.7$ and placed slits on many tens of galaxies. In principle, if the X-rays were associated with a structure at lower redshift, based on the implied X-ray luminosity of the detected X-ray source, we would expect much more than 10 galaxies within this radius of the X-ray centroid to have concurrent redshifts. We targeted emission line galaxies for their efficiency at yielding redshifts. These data revealed no new structure peaks along the line of sight. We therefore conclude that the X-ray source identified at the location of SpARCS1049 must be associated with SpARCS1049. 

\bibliography{sample63}{}
\bibliographystyle{aasjournal}



\end{document}